\documentclass[aps,prd,preprint,floats,tightenlines]{revtex4}

\newcommand{\beq}{\begin{equation}}
\newcommand{\eeq}{\end{equation}}

\begin{document}

\title{On cosmic natural selection}

\author{Alexander Vilenkin}

\address{
Institute of Cosmology, Department of Physics and Astronomy,\\ 
Tufts University, Medford, MA 02155, USA
}

\begin{abstract}

The rate of black hole formation can be increased by increasing the
value of the cosmological constant. This falsifies Smolin's conjecture
that the values of all constants of nature are adjusted to maximize
black hole production.

\end{abstract}

\maketitle

One of the fundamental questions in physics is why the constants of
nature have their observed values. This question becomes even more
intriguing if we observe that the constants appear to be tuned for the
existence of life. This can be explained in the context of multiverse
models, where the constants vary from one place in the universe to
another. The observed values are then determined partly by chance and
partly by anthropic selection (for an up to date discussion, see
\cite{Carr}).

An alternative explanation, suggested by Lee Smolin, is ``cosmic
natural selection'' \cite{Smolin1,Smolinbook,Smolin2}. Smolin assumes
that every time a black hole is formed, a new, causally disconnected
universe is created behind its horizon. The constants of nature in
this daughter universe are assumed to be slightly different from those
in the original one.  Smolin claims that this ``reproduction'' of
universes leads to an ensemble dominated by universes producing the
largest possible number of black holes. Furthermore, he claims that
the conditions maximizing the formation of massive stars turning into
black holes are about the same as the condition needed for the
evolution of carbon-based life. This explains the apparent fine-tuning
of the constants.

Smolin points out that his theory would be falsified if black hole
production were shown to increase when the constants of nature are
varied from their present values.  He has repeatedly challenged the
physics community to refute his theory and maintains that, despite
several attempts \cite{Ellis,Harrison,Silk,Susskind2}, it has not yet
been falsified.  Recently, there has been much discussion of Smolin's
ideas, especially in the popular press.  In this note, I will argue
that black hole production can be enhanced by an increase in the value
of the cosmological constant, thus falsifying Smolin's conjecture.

The simplest interpretation of the observed accelerated expansion of
the universe is that it is driven by a constant vacuum energy density,
which is about 3 times greater than the density of nonrelativistic
matter. Ordinary matter is being diluted, while the vacuum energy
density remains the same, and in another 10 billion years or so the
universe will be completely dominated by the vacuum. The following
evolution of the universe is accurately described by de Sitter
space. One might think that no new black holes will be formed in such
a universe, but this is not the case.

It has been shown by Gibbons and Hawking \cite{GH} that the state of
quantum fields in de Sitter space is similar to a thermal state with a
characteristic temperature
\beq
T_{GH}=H/2\pi,
\eeq
where 
\beq
H=(\Lambda/3)^{1/2}
\label{H}
\eeq
is the de Sitter expansion rate, $\Lambda$ is the cosmological
constant, which is related to the vacuum energy density as
$\Lambda=8\pi\rho_v$, and I am using Planck units in which the Planck
mass is $m_P =1$. (In these units, the present Hubble expansion rate
is $H\sim 10^{-61}$.) Quantum fluctuations of geometry will
result in occasional formation of black holes.  This is a quantum
tunneling process, which is mathematically described by an instanton
-- a solution of Euclideanized Einstein's equations.

One might be skeptical that we can say something definite about a
quantum process involving gravity, since we do not have a full theory
of quantum gravity. However, the instanton description of tunneling is
based on {\it semiclassical} gravity, which only requires the
knowledge of classical field equations and is generally believed to be
reliable.

The semiclassical nucleation rate of black holes per unit volume per
unit time, $\Gamma$, has been calculated by Ginsparg and Perry
\cite{Perry}, Chao \cite{Chao}, and by Bousso and Hawking
\cite{Bousso}. For black holes of mass $M\ll H^{-1}$, it is given by
\cite{Chao,Bousso} 
\beq
\Gamma \sim \exp(-M/T_{GH}), 
\label{Gamma}
\eeq
as might be expected. The rate decreases with $M$ and reaches the
minimum value for the largest black holes that can fit in de
Sitter space, having their Schwarzschild radius equal to the de Sitter
horizon. The mass of such black holes is $M\sim H^{-1}$, and their 
nucleation rate is \cite{Perry} 
\beq
\Gamma_{min} \sim\exp(-\pi/3H^2)=\exp(-\pi/\Lambda). 
\label{Gammamin}
\eeq

Since de Sitter space is eternal to the future, the total number of
created black holes is infinite. It is not clear then how the numbers
of black holes produced in different universes are to be
compared. This appears to be a serious difficulty, similar to the
measure problem of eternal inflation \cite{LLM,GSPVW}, but for the
time being it seems natural to assume that the maximum of probability
corresponds to the highest {\it rate} of black hole
production. According to Eqs.~(\ref{Gamma}),(\ref{Gammamin}), the rate
grows exponentially with $\Lambda$ and approaches one black hole per
Planck volume per Planck time as $\Lambda$ gets close to the Planck
scale ($\Lambda\sim 1$). At that point semiclassical gravity becomes
unreliable, and Eq.(\ref{Gamma}) cannot be extended to $\Lambda\gtrsim
1$.

If, for some reason, one does not trust semiclassical gravity, one can
argue on general grounds that quantum fluctuations resulting in a
local increase of energy density and leading to black hole formation
do not violate any conservation laws and should therefore have a
nonzero probability. Even if one insists that universes can be
created only in black holes resulting from stellar collapse, suitable
stars will pop out as quantum fluctuations in de Sitter space at a
nonzero rate (assuming that the radius of the star is much smaller
than the de Sitter horizon).  Quantum fluctuations generally get
stronger when $\Lambda$ is increased (since the effective de Sitter
temperature gets higher). Thus, black hole production should increase
with increase of $\Lambda$ \cite{foot1}.

For the observed value of $\Lambda$, the nucleation rate (\ref{Gamma})
is extremely small. With $M\gtrsim 1$, Eq.~(\ref{Gamma}) gives $\Gamma
\lesssim \exp(-10^{61})$ \cite{foot2}. And yet this process dominates
over any black holes that have been or will be produced by stellar
collapse, since the number of nucleated black holes is infinite. A
slight increase in $\Lambda$ would enhance the nucleation rate, thus
falsifying Smolin's conjecture.

One caveat is that the observed dark energy may not be a true
cosmological constant and may be gradually decreasing with time,
approaching zero or even a negative value. Then the de Sitter
expansion may stop before any black holes had a chance to
nucleate. But even if the true vacuum energy is constrained to be zero
or negative, a vacuum dominated stage can be achieved by varying the
parameters of the Higgs potential, so that a high-energy vacuum state
becomes metastable with a suffucuently long lifetime. A metastable
vacuum will decay through bubble formation, but if the bubble
nucleation rate satisfies
\beq
\Gamma_{bubble}\ll H^4,
\eeq
the volume occupied by the false vacuum will grow exponentially with
time \cite{Guth} and an infinite number of black holes will be
produced.  It is hard to imagine a realistic broken-symmetry theory of
elementary particles whose parameters cannot be adjusted to arrange a
metastable high-energy vacuum. An enormous number of such vacua is
suggested in string theory \cite{KKLM,Susskind}.

I would like to conclude with a general remark. One of the objections
Smolin raised against anthropic predictions is that they rely on the
principle of mediocrity, while there is no guarantee that our
civilization is typical \cite{Smolin2}. The same criticism, however,
can be applied to natural selection. The natural selection model does
not fix the constants precisely; it only gives a probability
distribution. The constants of nature may be far away from the peak of
the distribution in some rare universes, so no prediction can be made
unless one assumes that our universe is typical.  (This point has been
recently emphasized by Vaas \cite{Vaas}.)

I am grateful to Jaume Garriga for useful discussions. This
work was supported in part by the National Science Foundation.


\begin{thebibliography}{99}
\bibitem{foot1}
In the context of inflationary cosmology, bubbles of inflating false
vacuum will also be spontaneously produced at a nonzero rate
\cite{EWeinberg,recycling}. Each of these bubbles will become a site
of eternal inflation, producing an infinite number of black holes. But
once again, the bubble nucleation rate grows with the increase of the
cosmological constant, and thus the black hole production should grow
as well.
\bibitem{foot2}
Black holes of mass $M < 1$ have radius smaller than Planck length and
evaporate on a time scale shorter than Planck time. Such black holes
cannot be regarded as semiclassical objects, and we do not consired
them here.
\end{thebibliography}
\end{document}